\documentclass{emulateapj}

\usepackage{aas_macros}
\usepackage{epstopdf}
\usepackage{epsfig}
\usepackage{natbib}
\usepackage{float}
\usepackage{gensymb}
\usepackage{amsmath}
\usepackage{lipsum}% http://ctan.org/pkg/lipsum
\usepackage{times}
\usepackage{comment}
\usepackage{color}
\usepackage{multirow}
\usepackage{soul}
\usepackage{CJKutf8}            % Chinese name
\usepackage{xspace}

\definecolor{purple}{RGB}{160,32,240}

\definecolor{purple2}{RGB}{120,72,240}

\begin{document}

\title{The Anisotropic Circumgalactic Medium of Massive Early-Type Galaxies}

\author{Huanian Zhang \begin{CJK*}{UTF8}{gkai}(张华年)\altaffilmark{1,2}
\end{CJK*}}
\author{Dennis Zaritsky\altaffilmark{2}}
\altaffiltext{1}{Department of Astronomy, 
Huazhong University of Science and Technology, Wuhan, Hubei 430074, China; huanian@hust.edu.cn}
\altaffiltext{2}{Steward Observatory, University of Arizona, Tucson, AZ 85719, USA.}

\begin{abstract}
Using measurements of the [O {\small III}], H$\alpha$ and [N {\small II}] emission line fluxes originating in the cool (T $\sim10^4$ K) gas that populates the halos of massive early-type galaxies with stellar mass greater than $10^{10.4}$ M$_\odot$,
we explore the recent conjecture that active galactic nucleus (AGN) activity preferentially removes the circumgalactic medium (CGM) along the polar (minor-axis)  direction. We find deficits in the mean emission line flux of [O {\small III}] 
and H$\alpha$ (65 and 43\%, respectively) along the polar vs. planar directions, although due to the large uncertainties in these difficult measurements the results are of marginal statistical significance  (1.5$\sigma$). More robustly (97 to 99.9\% confidence depending on the statistical test), diagnostic line ratios show stronger AGN ionization signatures along the polar direction at small radii than at other angles or radii.  Our results are consistent with the conjecture of an anisotropic CGM in massive, early type galaxies, suggested on independent grounds, that is tied to AGN activity and begin to show the potential of CGM mapping using emission lines.
\end{abstract}


\section{Introduction}

\cite{Ignacio2021} identified a statistically significant azimuthal variation in the star-formation rates of  satellite galaxies of massive early type galaxies.
They proposed that AGN-powered outflows modify the circumgalacitc medium (CGM) along the host's minor axis, reducing the ram pressure on orbiting satellites whose orbits are more closely aligned with that axis and helping preserve their star formation rates relative to that of those satellites whose orbits are more closely aligned with the host's major axis. We test this conjecture by measuring the properties of the CGM as a function of azimuthal angle for a comparable set of massive early-type galaxies. 

Studies of the CGM have primarily relied on measurements of absorption lines in the spectra of bright background objects \citep[e.g.][]{steidel2010,Chen2010,menard2011,bordoloi2011,zhu2013a,zhu2013b,Werk2013,Johnson2013,Johnson2014,Werk2014,Johnson2015,werk16,croft2016,croft2018,prochaska2017,Cai2017,Johnson2017,Chen2010,Chen2017a,lan2018,joshi2018,Chen2019,Zahedy2019,Dutta2020,Zheng2020,Haislmaier2021,CGMsquare2021,Norris2021,Qu2022}. This is a rich field of study, in some cases with a particular focus on massive galaxies like those of concern here \cite[e.g.,][]{Chen2017b,Zahedy2020}, that has yielded many insights into the galactic baryon cycle \citep{Donahue2022}.  Unfortunately, the total number of measured sightlines remains relatively small, making it statistically difficult to compare CGM properties among specific, narrow subsets of the available sample.

A growing set of complementary studies is now focusing on measurements of emission lines. Optical emission lines from the CGM provide an opportunity to explore the cool ($T \sim 10^4$ K) phase of the CGM, but are challenging to measure \citep{zhang2016}. In the local universe, H$\alpha$ has been detected in the CGM of {\sl individual} galaxies only when the systems are extreme \citep[such as in the starburst/merger NGC 6240;][]{Yoshida2016}. 
\cite{zhang2016} presented the first detection of %the emission line
%flux of 
H$\alpha$ and [N{ \small II}] $\lambda$6583, from low redshift, normal galaxies extending out to a projected radius of $\sim$ 100 kpc by stacking a sample of millions of sightlines from the Sloan Digital Sky Survey \citep[SDSS DR12;][] {SDSS12}. Because every galaxy, in principle, has a CGM that is emitting, large statistical samples are straightforwardly compiled and specific questions, such as that regarding the azimuthal properties of the CGM, can be addressed using galaxy ensembles. 

Building on that first H$\alpha$  result using SDSS spectra, subsequent studies have characterized the line-emitting, cool CGM within 50 kpc  or one-quarter of the virial radius, $0.25R_{\rm vir}$, in low redshift galaxies \citep[][hereafter, Papers I, II, III, IV, V, VI, VII]{zhang2016,zhang2018a,Zhang2018b, Zhang2019, Zhang2020a, Zhang2020b, Zhang2021}. 
As in most of those studies, we restrict ourselves here to radii interior to 50 kpc or $0.25R_{\rm vir}$ because the measured emission beyond this radius is strongly contaminated by emission from nearby associated galaxies (Paper II). 

From among the previous studies, the most relevant here is Paper III, which presents a study of the physical properties of the CGM based on diagnostic line ratios.
Line ratios, like those used in the BPT diagram \citep{bpt}, provide guidance on the ionizing source of the gas. The use of 
such line ratios
has become common in the study of the central galaxies, particularly to distinguish between the two expected dominant sources of ionization, star formation and active galactic nuclei (AGN) \citep[{e.g.}][]{vo,kewley,kauffmann_agn}. 
In Paper III, we  found that lower mass galaxies, $M_* < 10^{10.4}$ M$_\odot$, have halo gas that is ionized by softer sources, similar to that found in star forming regions (H {\small II} regions), while higher mass galaxies, $M_* > 10^{10.4}$ M$_\odot$, have halo gas that is ionized by harder sources, similar to that found in AGN-hosting galaxies or in shocked regions. Here we use the same diagnostic tool, but examine the CGM behavior at different position angles, or azimuthal angles, around massive galaxies and at different radii.

This paper is organized as follows. In \S\ref{sec:dataAna} we present the data analysis, including sample selection and a reprise of  the basics of our technique. In \S\ref{sec:results} we present our measurements and identify any statistically significant differences we find as a function of azimuth and radius. In \S\ref{sec:discussion} we 
discuss implications of the results in the context of the \cite{Ignacio2021} scenario. In \S\ref{sec:sum}, we summarize and conclude.
Throughout this paper, we adopt a $\Lambda$CDM cosmology with parameters
$\Omega_m$ = 0.3, $\Omega_\Lambda =$ 0.7, $\Omega_k$ = 0 and the dimensionless Hubble constant $h = $ 0.7 \citep[cf.][]{riess,Planck2018}.

\section{Data Analysis}
\label{sec:dataAna}

\begin{figure}[htbp]
\begin{center}
\includegraphics[width = 0.48 \textwidth]{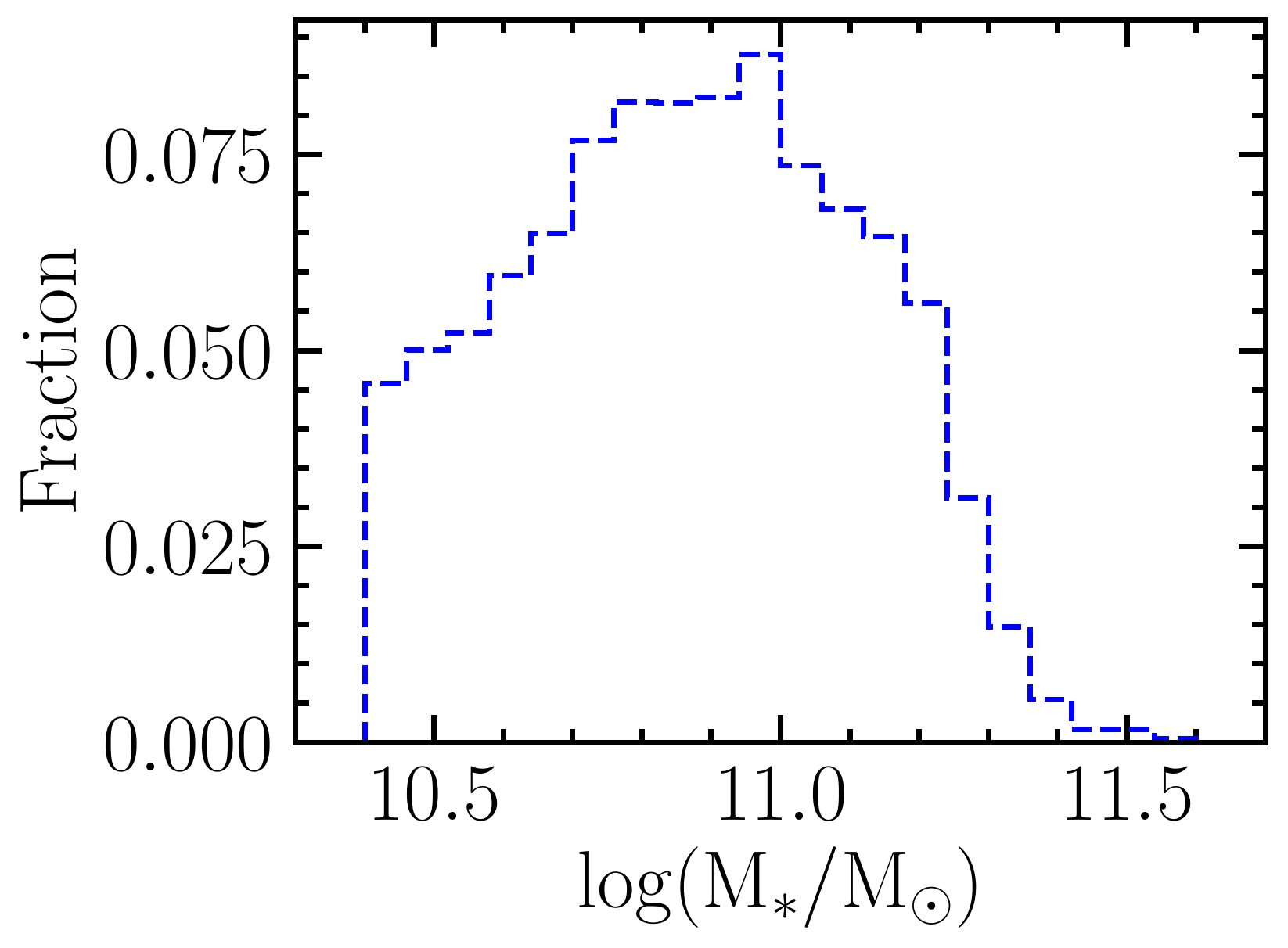}
\end{center}
\caption{The distribution of stellar mass for the primary galaxy sample. We require M$_* > 10^{10.4}$ M$_\odot$ to match
the stellar mass where we found the CGM to have AGN-like properties (Paper III) and to approximately match the \cite{Ignacio2021} study. }  
\label{fig:sm}
\end{figure}

We follow the approach developed in Papers I through VII by selecting galaxies that meet our standard criteria in redshift ($0.02 < z < 0.2$),  half light radius ($1.5 < R_{50}/{\rm kpc} < 10$),  and  $r-$band luminosity ($10^{9.5} < L_r /L_\odot < 10^{11}$) and add additional criteria to produce the closest match to the \cite{Ignacio2021} sample.   For the primary galaxies, we extract measurements of the position angle, the S\'ersic index ($n$), the ellipticity ($e$) and $r$-band absolute magnitude ($M_r$) from \citet{simard}. This selection limits
the primary sample to galaxies from the 7th major data release of SDSS (DR7). We extract measurements of stellar mass (M$_*$) from  \cite{Kauffmann2003a,Kauffmann2003b} and \cite{Gallazzi}, and star formation rates (SFR) from the MPA-JHU catalog \citep{Brinchmann}.  The SFR estimates are aperture corrected to account for the light outside the SDSS/eBOSS fiber aperture (2 arcsecond), which only collects $\sim$ 1/3 of the total light for a typical galaxy at the median redshift of the
survey \citep[for details see][]{Brinchmann}. We require the primary galaxies to be early type, defined as having S\'ersic index $n > 2.5$, and to have a stellar mass such that M$_* > 10^{10.4}$ M$_\odot$,
which matches the stellar mass where we found the CGM to have AGN-like properties (Paper III) and which roughly corresponds to a halo mass of $10^{12}$ M$_\odot$ \citep{Behroozi2010,Behroozi2019}, as selected by \cite{Ignacio2021}.  We present the distribution of the galaxy stellar masses in Figure \ref{fig:sm}. The mean and median mass of the primary galaxy sample are consistent, $\sim 10^{10.87}$ M$_\odot$. Finally, we also define a minimum ellipticity criterion ($e > 0.25$) to ensure that the position angle of the major axis is well-defined. We will discuss the impact of the limit on $e$ on our results in \S\ref{sec:results}.

We use spectra from SDSS/eBOSS DR16 \citep{SDSSDR16} for all of the sightlines projected within a specified range of scaled projected radii for each of our primary galaxies. As described in Paper V,  we use scaled, projected radii, $r_s$, to account for the range in primary galaxy sizes. We define  $r_s$ as the ratio between the physical projected separation and the virial radius of the primary galaxy ($r_s  \equiv r_p/r_{\rm vir}$). To estimate the virial radius of the primary galaxy, we use the scaling relation between the luminosity and the virial radius obtained by fitting a high order polynomial to the results drawn from the UniverseMachine \citep{Behroozi2019}. As discussed in more detail in Paper III, 
we typically set a physical lower limit on the projected radius (10 kpc) for our sightlines to mitigate possible contamination of the spectra by the central galaxy, but in this study that criterion is superseded by our $r_s$ lower limit of 0.05, which roughly corresponds to $\sim$ 15 kpc for galaxies with M$_* > 10^{10.4}$ M$_\odot$.

\begin{figure}[htbp]
\begin{center}
\includegraphics[width = 0.48 \textwidth]{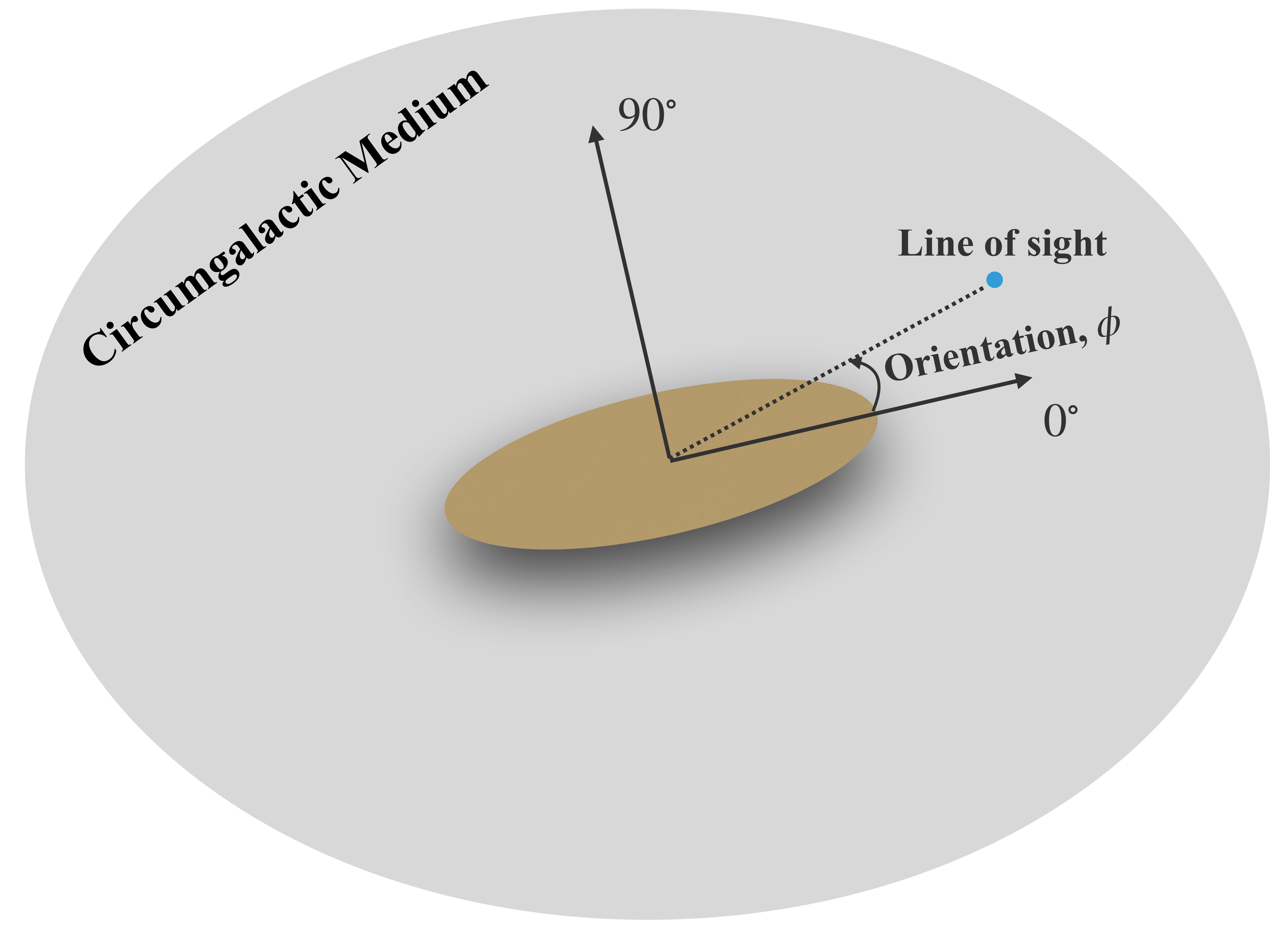}
\includegraphics[width = 0.48 \textwidth]{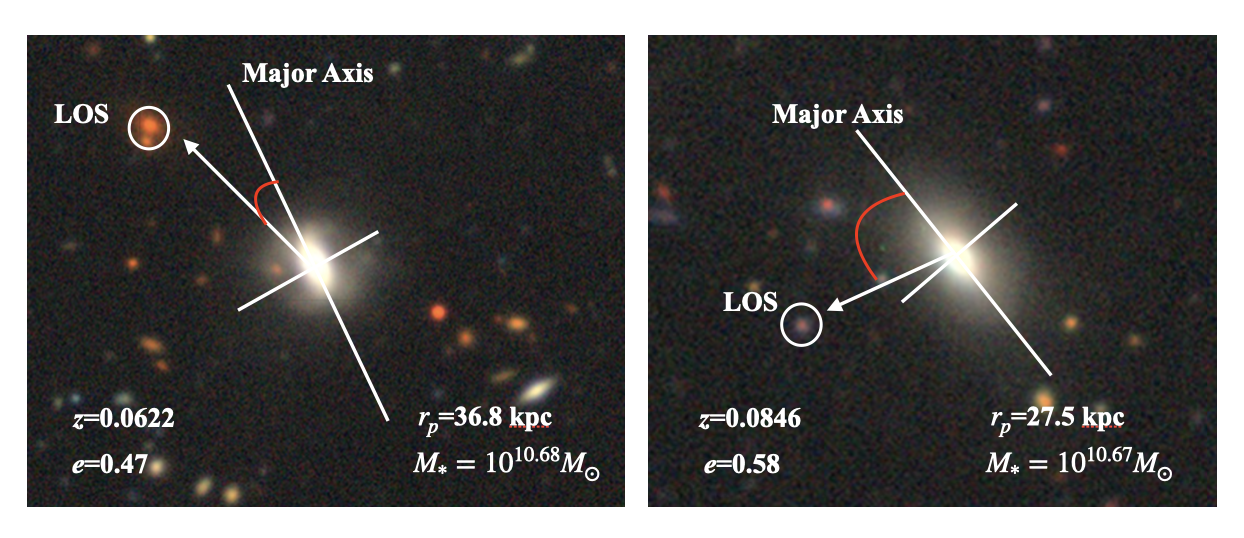}
\end{center}
\caption{In the top panel we show a cartoon of  the orientation of the sightline with respect to the target galaxy. A value of $\phi = 0^\circ$ represents the alignment along the major axis of the target galaxy, while $90^\circ$ along the minor axis. In the bottom panels we present two galaxies from our sample, with images drawn from  from the DESI Legacy Imaging Surveys \citep{Arjun2019}, and the sightline orientation angles ($\phi$) indicated with red arcs. The orientation angle of the example on the left is 18.11$^\circ$. It is 81.67$^\circ$ for the example on the right. The target galaxy on the left is located at (RA, DEC) = (1.61904$^\circ$, +0.48484$^\circ$) and has $z = 0.0622$, $e = 0.47$ and M$_* = 10^{10.68}$ M$_\odot$. The projected radius to the sightline is 36.8 kpc. The target galaxy on the right is located at (RA, DEC) = (18.2994$^\circ$, +0.6965$^\circ$) and has $z = 0.0846$, $e=0.58$ and M$_* = 10^{10.67}$ M$_\odot$. The projected radius to the sightline is 27.5 kpc.}  
\label{fig:cgm_orien}
\end{figure}

To probe the azimuthal distribution of the CGM, we calculate the orientation angle, $\phi$, of each sightline with respect to the major axis of each target galaxy.
Specifically, we first calculate the angle on the sky made by the line connecting the center of the primary galaxy and the position of the sightline,
PA$_{1,2}$, using
\begin{equation}
\label{eq:pa}
\tan{(\rm PA_{1,2})} = \frac{\sin(\alpha_1 - \alpha_2)}{\cos \delta_2 \cdot  \tan \delta_1 - \sin \delta_2 \cdot \cos(\alpha_1 - \alpha_2)}
\end{equation}
where $(\alpha_1, \delta_1)$ and $(\alpha_2, \delta_2)$ are the right ascension and declination of the primary galaxy and the sightline. 
We then calculate the difference between PA$_{1,2}$ and the major axis position angle of the target galaxy, restricting the difference to the range of 0$^\circ$ to 90$^\circ$, where 0$^{\circ}$ corresponds to the sightline lying along the major axis and $90^\circ$ along the minor axis, and refer to the angle as $\phi$ (see Figure \ref{fig:cgm_orien}).

Our procedure for  processing the sightline spectra follows from our previous papers.
For each spectrum, we fit and subtract a 10th order polynomial to a 300 \AA\ wide section surrounding the observed wavelength of H$\alpha$ at the primary galaxy redshift to remove the continuum.  
We then measure the residual H$\alpha$ flux within a velocity window centered on the recessional velocity of the primary galaxy.  We adopt velocity windows of $\pm 330$ km s$^{-1}$ and $\pm 450$ km s$^{-1}$ for galaxy stellar masses in the range of $10^{10.4} < {\rm M_*/M_\odot} \le 10^{11}$ and ${\rm M_*/M_\odot} > 10^{11}$, respectively. See Paper V for more details.  We only analyze sightlines where the continuum level is $< $ 3 $\times$ $10^{-17}$ erg cm$^{-2}$ s$^{-1}$ \AA$^{-1}$ to limit the noise introduced by the actual SDSS spectral target and require that the measured emission line flux be within 3$\sigma$ of the mean of the whole sample to remove spectra of interloping strong emitters such as satellite galaxies.  We apply the same procedures and criteria for the [O {\small III}]$\lambda 5007$ and [N {\small II}]$\lambda 6583$ emission lines except that the continuum level cut for [O {\small III}]$\lambda 5007$ is $< $ 2.0 $\times$ $10^{-17}$ erg cm$^{-2}$ s$^{-1}$ \AA$^{-1}$. We have confirmed that using the mean or median of the resulting set of flux measurements in our subsequent analysis produces consistent results. We present the results of mean flux here.

We estimate the uncertainties in the mean flux values by randomly selecting half of the individual spectra in the relevant subsample, calculating the mean emission line flux, and repeating the process 1000 times to establish the distribution of measurements from which we quote the values corresponding to the 16.5 and 83.5 percentiles as the uncertainty range. We compensate for using only half the sample in each measurement by dividing the resulting $1\sigma$ estimated uncertainties by a factor of $\sqrt{2}$.

\section{Results}
\label{sec:results}
Our primary interest here is to examine the suggestion by \cite{Ignacio2021} that there are physical differences in the CGM of massive early type galaxies between the polar and planar directions. To do this
we separate our measurements into two bins based on $\phi$, 
$0^\circ \le \phi  < 45^\circ$  and
$45^\circ \le \phi < 90^\circ$, which constrain the CGM properties
along the major and minor axes of the target galaxy, respectively. 
Although \cite{Ignacio2021} posited that the influence of the AGN activity reaches beyond the virial radius, we constrain our examination to projected radius between 0.05$r_{\rm vir}$ and 0.25$r_{\rm vir}$ because of contamination from emission line flux arising in nearby, associated halos at larger radii (Paper II).

\begin{figure}[ht]
\begin{center}
\includegraphics[width = 0.48 \textwidth]{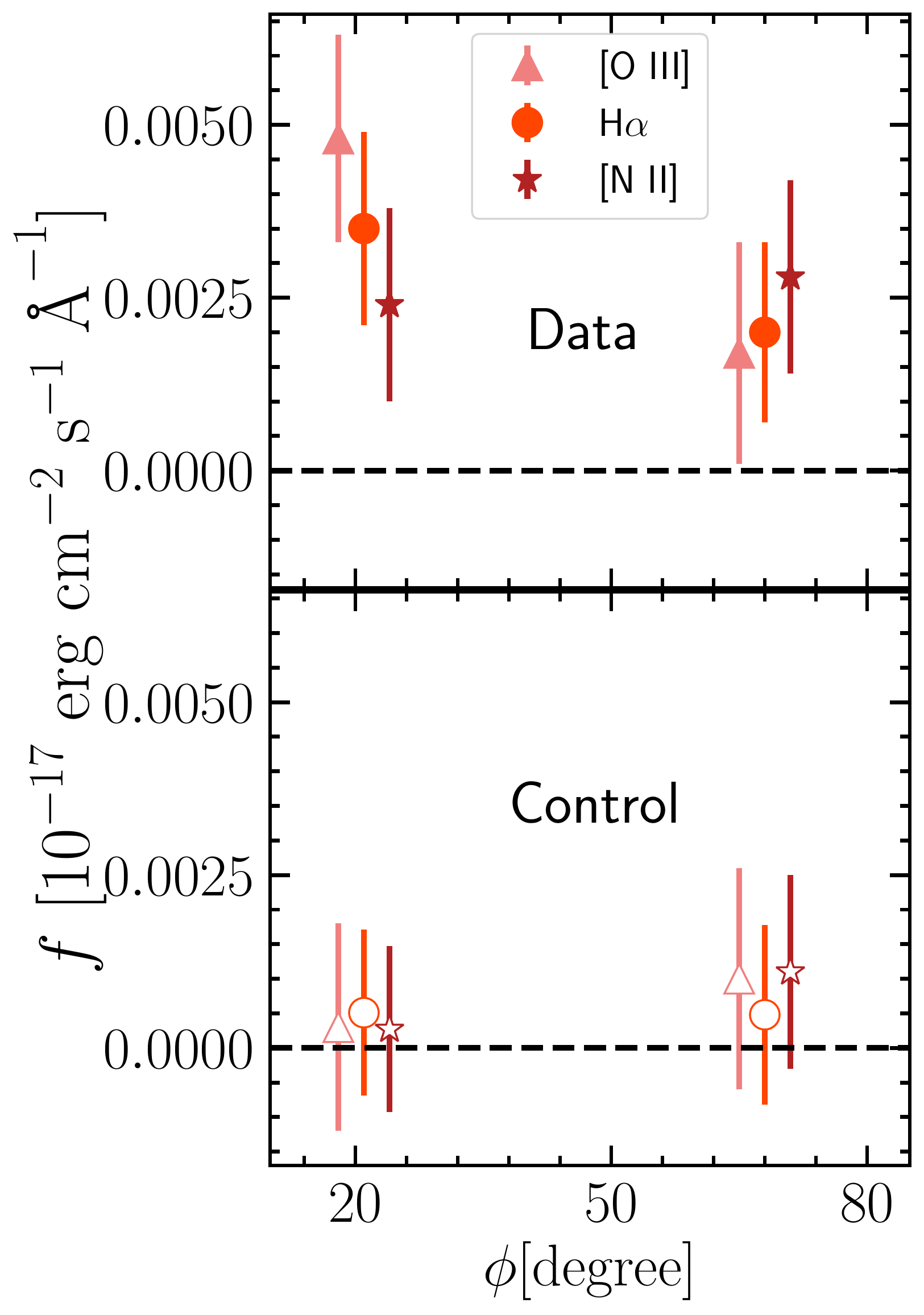}
\end{center}
\caption{The [O {\tiny III}], H$\alpha$ and [N {\tiny II}] emission line flux for data sample (solid shapes) and control sample (open shapes) as a function of azimuthal angle. The data points are separated in horizontal direction for better visualization. The black dashed line indicates the zero flux for better visualization.  }  
\label{fig:flux_angle}
\end{figure}

In our spectral stacks, we detect three emissions lines,
[O {\small III}]$\lambda$5007, H$\alpha$ and [N {\small II}]$\lambda$6583 (Table \ref{tab:flux_angle} and Figure \ref{fig:flux_angle}). There is a fractionally large, but statistically marginal, drop in the [O {\small III}] and H$\alpha$ fluxes when moving from the primary galaxy's major axis toward its minor axis.
For example, the decline in [O {\small III}] flux is $0.0031\pm0.0022$ in units of 10$^{-17}$ erg cm$^{-2}$ s$^{-1}$ \AA, which corresponds to a flux drop of 65\% but is only a 1.5$\sigma$ detection. The apparent drop in flux is consistent in the sense expected from the scenario presented by \cite{Ignacio2021}, but is not yet statistically convincing. The result of combining all of the line fluxes and comparing in azimuth also yields only a 1.5$\sigma$ detection.

\begin{deluxetable}{cccc}
\tablewidth{0pt}
\tablecaption{The stacked [O {\tiny II}], H$\alpha$ and [N {\tiny II}] emission flux vs azimuthal angle.}
\tablehead{  \colhead{Line} & \colhead{$\langle \phi \rangle$} & \colhead{N} & \colhead{$f$}\\
 &[$^\circ$]&&\colhead{
 [$10^{-17}$\,erg\,cm$^{-2}$\,s$^{-1}$\,\AA$^{-1}$]}
 }
\startdata
\multirow{2}{*}{[O {\tiny III}]} & 21 & 2966 & $0.0048 \pm 0.0015$ \\  
& 68 & 2982 & $0.0017 \pm 0.0016$  \\ \\
\multirow{2}{*}{H$\alpha$} & 21 & 3098 & $0.0035 \pm 0.0014$  \\ 
& 68 & 3098 & $0.0020 \pm 0.0013$  \\ \\
\multirow{2}{*}{[N {\tiny II}]} & 21 & 3046 & $0.0024 \pm 0.0014$ \\
& 68 & 3081 & $0.0028 \pm 0.0014$  \\
 \enddata
\label{tab:flux_angle}
\end{deluxetable}

To confirm our ability to make such  measurements and estimate uncertainties, we construct a control sample as in Paper IV, in which we `move' each primary galaxies to a blank sky position, assign a random intrinsic position angle to the primary galaxy, and redo the full analysis. Clearly in such cases we should find no emission flux and no dependence on azimuth. In Figure \ref{fig:flux_angle}, we include the results from this control experiment, for which the results are indeed consistent with zero flux and no dependence with orientation angle to within the estimated uncertainties.

Next, we apply a test that is independent of binning in $\phi$ to further investigate the statistical significance of any difference the H$\alpha$ emission flux as a function of azimuth. We construct a vector that corresponds to each individual line of sight measurement that has a magnitude corresponding to the measured H$\alpha$ flux and orientation $\phi$. The vector sum for all of the lines of sight with $0.05 < r_s < 0.125$ lies along the primary galaxy major axis with offset angle of only 3.8$^\circ$, suggesting a strong excess of flux along the major axis. We randomize the fluxes and position angles in 10000 trails and find that only $\sim$ 4.6\% of the time does the sum align within  3.8$^\circ$ of the major axis. We apply the same test to the [O {\small III}] within $0.125 < r_s < 0.25$ and find that the sum is offset from the primary's major axis by $\sim$ 14.5$^\circ$, which occurs in only $\sim$ 3.9\% of the corresponding 10000 random trials. Both results support the previous findings that the emission line fluxes decline toward the polar axis.

\begin{figure}[htbp]
\begin{center}
\includegraphics[width = 0.48 \textwidth]{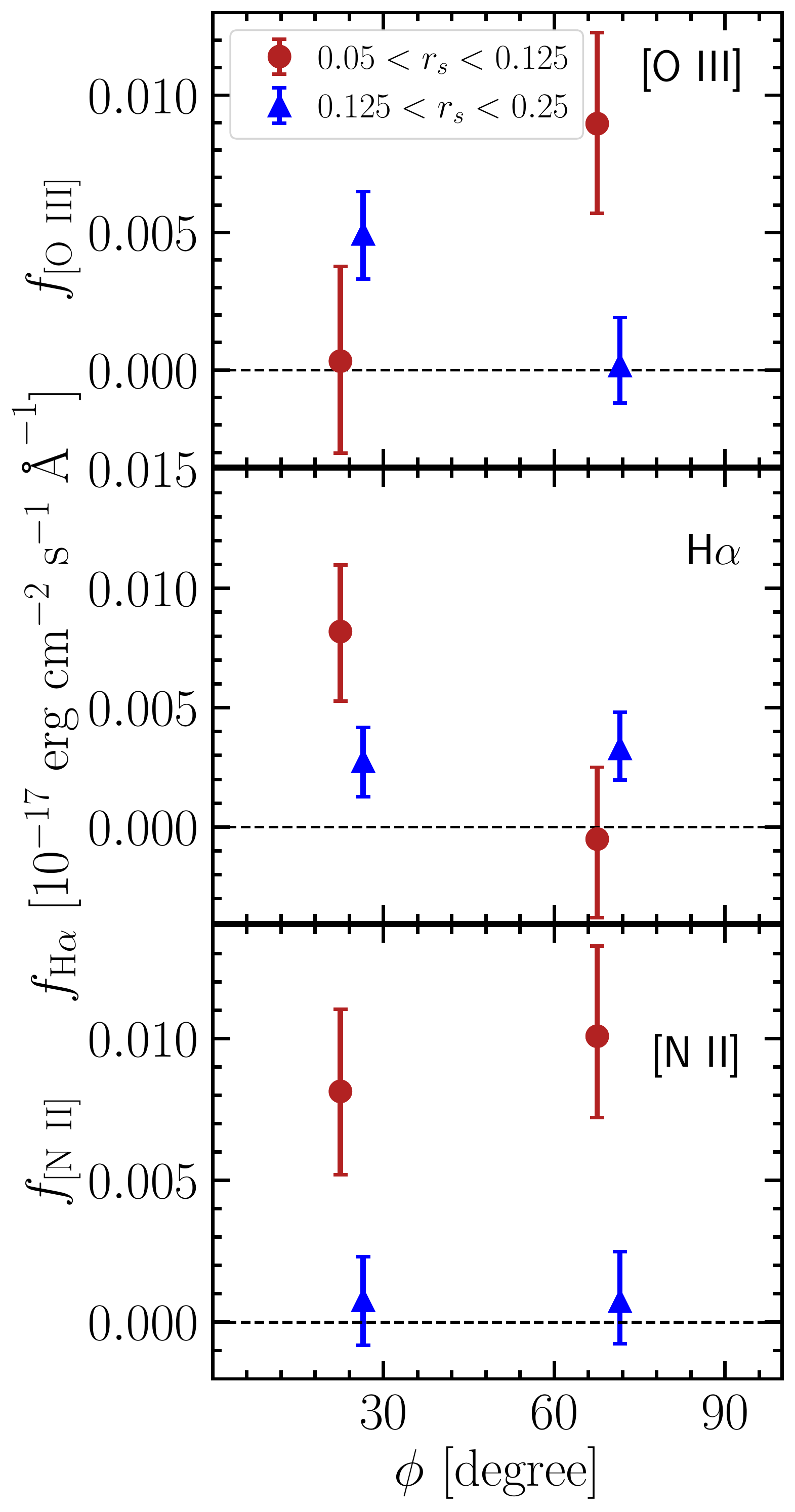}
\end{center}
\caption{The [O {\tiny III}] (top), H$\alpha$ (middle) and [N {\tiny II}] (bottom) emission line flux as a function of azimuthal angle. The red dot represents the emission flux at the inner radii of $0.05 < r_s < 0.125$ and the blue triangle is for the outer radii of $0.125 < r_s < 0.25$. The horizontal dashed line indicates the zero flux for better visualization.}  
\label{fig:flux_rs}
\end{figure}

Finally, to explore if azimuthal differences are more statistically significant over certain radial ranges, we divide the data into two radial bins.  For each of the same two azimuthal bins, we now present average measurements in each of two $r_s$ bins, $0.05 < r_s < 0.125$ and $0.125 < r_s < 0.25$ (Table \ref{tab:flux_rs}  and Figure \ref{fig:flux_rs}). The differences we find are 1) an azimuthal decline in H$\alpha$ flux in the inner radial bin (2.0 $\sigma$) as one progresses from sightlines along the major axis to those along the minor axis, 2) a radial decline in [O {\small III}] flux along the minor axis (2.0 $\sigma$), and 3) a radial decline in [N {\small II}] flux that is independent of azimuth (2.2$\sigma$ and 2.8$\sigma$, along the major and minor axes, respectively). Again differences exist but are statistically marginal.

The results so far hint at azimuthal and radial variations in CGM properties, but are often of marginal statistical significance and, at least superficially, can sometimes appear to go in opposite directions, for example with the H$\alpha$ flux decreasing toward the polar direction and the [O {\small III}] flux increasing toward the polar direction in the innermost radial bin. If these variations are real, they suggest a more complex behavior than a simple removal of gas along the minor axis. Given that the cause of these variations is posited to be related to nuclear activity in the galaxy, we proceed to examine a line diagnostic ratio that is widely used to quantify the nature of the ionizing source.

\begin{deluxetable}{cccrr}
\tablewidth{0pt}
\tablecaption{The stacked [O III], H$\alpha$ and [N II] emission flux vs azimuthal angle and radius}
\tablehead{  \colhead{Line} & \colhead{$\langle \phi \rangle$} & \colhead{$r_s$}\tablenotemark{a} & \colhead{N} & \colhead{$f$}\\
 &[$^\circ$]&&&\colhead{
 [$10^{-17}$\,erg\,cm$^{-2}$\,s$^{-1}$\,\AA$^{-1}$]}
 }
\startdata
\multirow{4}{*}{[O {\tiny III}]} & \multirow{2}{*}{21} & 0.09 & 642 & $0.0003 \pm 0.0033$ \\
& & 0.19 & 2324 & $0.0052 \pm 0.0017$ \\
& \multirow{2}{*}{68} & 0.09 & 651 & $0.0089 \pm 0.0032$ \\
& & 0.19 & 2331 & $0.0001 \pm 0.0014$ \\ \\
\multirow{4}{*}{H$\alpha$} & \multirow{2}{*}{21} & 0.09 & 688 & $0.0082 \pm 0.0029$ \\
& & 0.19 & 2410 & $0.0025 \pm 0.0015$ \\
& \multirow{2}{*}{68} & 0.09 & 703 & $-0.0005 \pm 0.0028$ \\
& & 0.19 & 2395 & $0.0032 \pm 0.0014$ \\ \\
\multirow{4}{*}{[N {\tiny II}]} & \multirow{2}{*}{21} & 0.09 & 689 & $0.0081 \pm 0.0029$ \\
& & 0.19 & 2357 & $0.0008 \pm 0.0015$ \\
& \multirow{2}{*}{68} & 0.09 & 694 & $0.0100 \pm 0.0029$ \\
& & 0.19 & 2387 & $0.0007 \pm 0.0015$ \\
 \enddata
\label{tab:flux_rs}
\tablenotetext{a}{$r_s$ is  the  ratio  between  the  physical projected separation and the virial radius of the primary galaxy, $r_s \equiv r_p / r_{\rm vir}$.}
\end{deluxetable}

\begin{figure}[htbp]
\begin{center}
\includegraphics[width = 0.48 \textwidth]{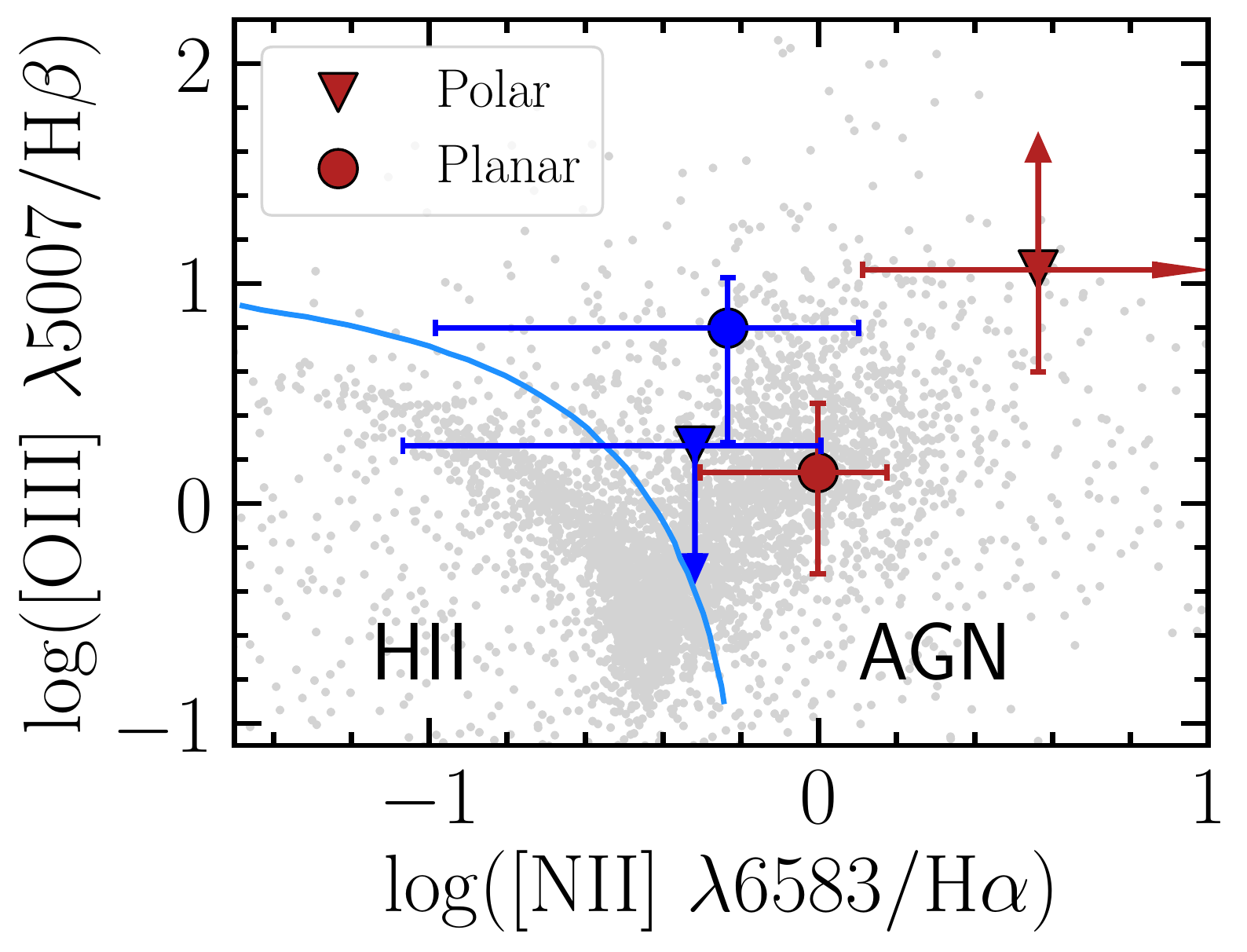}
\end{center}
\caption{The BPT emission line ratios for two azimuthal and two radial bins. The inner bins, $0.05 < r_s < 0.125$,  are represented in red, while the outer, $0.125 < r_s < 0.25$, are in blue. The symbols designate the azimuthal bins as given in the legend. The blue curve shows the boundary between the H{\small II} and AGN regions of the diagram and the light grey points represent individual SDSS galaxies.}
\label{fig:bpt}
\end{figure}

As we have discussed previously, a standard line diagnostic diagram is referred to as the BPT diagram \citep{bpt}, which compares [N{\small II}]/H$\alpha$ and [O{\small III}]/H$\beta$.
One challenge in calculating this set of line ratios is that we do not have the S/N to detect H$\beta$ in the spectral stacks. As we did in Paper III when calculating the BPT line ratio,
we adopt H$\beta$/H$\alpha = 0.3$,  a rough value consistent with the non-detection for our entire sample and with theoretical expectations. 
A second challenge, is that in some cases we have average flux measurement that are formally consistent with zero. In these cases, we use the $1\sigma$ upper limit when calculating the related line ratio and quote the result as the corresponding limit on the line ratio.
We present the BPT line ratios, or limits, 
for the two azimuthal and the two radial bins in Figure \ref{fig:bpt}.

The line ratios presented in Figure \ref{fig:bpt} are consistent with ionization of the CGM by AGN, as expected for massive galaxies with stellar mass $> 10^{10.4}$ M$_\odot$ (Paper III). Although the uncertainties are large, the results in one bin (the inner bin along the minor axis) appear to be quite different than those of the other bins. Beyond the statistical significance of the difference, which we discuss next, we note that
the offset is in the expected sense in that the inner CGM along the minor axis most strongly  reflects the properties of AGN-ionized gas. 

To quantitatively assess the significance of this result, we reconstruct binned averages, randomly drawing with replacement, samples to match the number of sightlines included in the inner, polar subsample from data in the three other bins.  The fraction of the 10,000 reconstituted samples where the result is at least as far toward the upper right of Figure \ref{fig:bpt} as the actual measurement is only 3\%. This test suggests that the inner, polar set of sightlines are different than all other sightlines with 97\% confidence. However, this approach combines sightlines across radii. If there are radial differences among sightlines, then we should only compare between inner polar and planar sightlines. When drawing solely from the inner, planar bin, with replacement, and repeating the test we find that inner, polar bin differs from the inner, planar bin with 99.7\% confidence. 

One concern here is that we could be mislead because the tests just described are {\sl a  posteori} statistical tests. To address this concern, we revisit our control sample and apply the same test. We repeat the analysis defining each of the four bins as the `different' bin and drawing comparison samples from the other three bins. In no combination is the sole bin found to be different than the other bins with greater than 1$\sigma$ confidence. Additionally, for 1,000 random draws of a sample of the size of original inner, polar sample of sightlines, only 0.1\% are offset as far or farther toward the upper right of the BPT diagram as the actual inner, polar sample. These tests, plus the qualitative nature of the finding (that the inner, polar bin is the one that shows the strongest AGN features), provide additional confidence in the finding.

\subsection{Dependence on galaxy properties}

As introduced in \S\ref{sec:dataAna}, we set a minimum ellipticity criterion for the primaries to ensure that the position angles are robustly measured. Although the need for such a criterion is evident, the value chosen is somewhat arbitrary but is an attempt to achieve a balance between having a larger primary sample that perhaps includes more for which the position angle is uncertain and a smaller sample with better defined position angles. It is not {\sl a priori} known which choice may result in the most statistically significant results. Applying a more conservative cut ($e > 0.4$) results in a sample that is half the size but a BPT ratio result that is nearly as significant as that described previously (99.3\% confidence vs. 99.7\%). On the other hand, a more generous cut ($e > 0.1$) increases the sample by about 50\% but the significance of the result drops to 90.6\%. Our initial choice of $e > 0.25$ strikes a balance and other choices do nor appear to provide more statistically significant results.

In Papers II/III/IV, we found that the emission line flux from the CGM correlates with the stellar mass, SFR, and morphology of the primary galaxy. As such, it is always a danger when comparing samples selected for one reason that any observed differences might instead originate from these differences. Although we cannot envision reasons why selecting on azimuth or radius of the sightlines would connect to the properties of the primary galaxy, we nevertheless test that the primaries in the various subsamples are all similar. The samples are indeed nearly indistinguishable, with mean stellar masses for the four samples ranging between 
$10^{10.90}$ M$_\odot$ and $10^{10.93}$ M$_\odot$, the S\'ersic $n$ indices between 5.01 and 5.05, and the SFRs between 0.36 and 0.41 M$_\odot$/yr. Based on previous results, Papers II/III/IV, variations within these ranges are not expected to lead to detectable emission line differences.

\subsection{Discussion}
\label{sec:discussion}

We find that the inner CGM is indeed different along the minor axis than along the major axis. Furthermore, this difference is in line with the hypothesis that the central AGN has a stronger influence in this direction. However, the emerging picture for the observed emission line behavior hints at something more complex than the simple interpretation posited by \cite{Ignacio2021} to explain their results, a lower density CGM along the minor axis.  Consider that in Table \ref{tab:flux_rs}  and Figure \ref{fig:flux_rs} the H$\alpha$ flux drops by more than a factor of 2 in going from the planar to polar orientations, naively suggesting a commensurate density change in the expected direction. However, the trend in [O {\small III}] flux is in the opposite sense (Table \ref{tab:flux_rs} and Figure \ref{fig:flux_rs}) suggesting that there are other differences at play beyond simply less gas. In retrospect, it is evident that
the influence of the AGN could not only lead to a lower density, but also have other effects such as in the mean metallicity of the gas \citep{Schaefer2020,Bao2021} and the ionizing spectrum. In fact, our most statistically significant result regards the nature of the ionizing spectrum. Untangling these different aspects is a challenge beyond our simple azimuthally-dependent analysis \citep[e.g., see the complications in the metallicity analysis of absorption line systems;][]{Gibson2022}. 

 Evidence for asymmetries in the CGM is also available from absorption line studies. 
\cite{Huang2016} studied the CGM of the  luminous red galaxies (LRGs) using SDSS LRG-QSO pairs. They found no strong dependence of Mg {\small II} covering fraction for either passive or [O {\small II}]-emitting LRGs at projected radii $>$ 50 kpc, but found a modest enhancement ($\sim 50$\%) of Mg {\small II} absorption closer to the major axis of [O {\small II}]-emitting LRGs at projected radii $<$ 50 kpc that decreases toward the minor axis. Our finding is qualitatively consistent with those findings.

Interpreting results is particularly fraught when stacking tens of thousands of spectra. The CGM in each galaxy is a multi-phase, multi-ionization-state and  geometrically-complex structure \citep{CGM2017}. It is not evident how those add up to produce a mean line flux. In broad strokes, the results of simple models are plausible (Paper VII), but in assessing the detailed interplay between AGN and CGM, in what is perhaps the average of a heterogeneous sample, such models are unlikely to be illuminating. Instead, we await the detection of emission lines in individual, normal, low-redshift galaxies. This is a challenging observation for current state-of-the-art facilities, but should become routine with the next generation of larger ground based telescopes. For the warm CGM such observations may come relatively sooner using ultraviolet emission lines observed with satellites like {\sl Aspera} \citep{aspera}, while for the hot CGM they may eventually be available from X-ray facilities such as the Hot Universe Baryon Surveyor \citep[HUBS,][]{HUBS2020}.

\section{Summary}
\label{sec:sum}

To test whether AGN activity affects the circumgalactic medium  anisotropically, we applied a methodology developed to measure the emission line fluxes from the cool ($T\sim 10^4$K) CGM (Paper I).
We selected primary galaxies with S\'ersic index $n > 2.5$,  ellipticity $e > 0.25$ and   stellar mass M$_* > 10^{10.4}$ M$_\odot$, which roughly corresponds to a halo mass of $10^{12}$ M$_\odot$, to closely match the galaxy sample studied by \cite{Ignacio2021}. We calculated the relative position angle between each sightline and the associated central galaxy major axis and combined the H$\alpha$, [O {\small III}], and [N {\small II}] fluxes to search for trends 
as a function of scaled projected radii and azimuthal angle.  

The emission line flux of [O {\small III}] and H$\alpha$ have a large ($\sim$ 65\% and $\sim$43\%, respectively) drop from the major axis to the minor axis of the central galaxy, consistent with the hypothesis that the CGM has a lower density in the polar direction \citep{Ignacio2021}, but due to large uncertainties the results are statistically marginal ($\sim$ 1.5$\sigma$). Further investigation of the azimuthal flux behavior in two different radial bins shows more complex behavior than can be attributed to merely a difference in gas density in the polar vs. planar directions, but these results are also statistically marginal.

Because the conjecture has nuclear activity affecting the CGM, we explore the BPT diagnostic line ratio for the CGM along the major and minor axis of the central galaxy. We adopt a fixed value of H$\beta$/H$\alpha$ because of the non-detection of H$\beta$ in our spectra and quote limits for the logarithm of the ratios when either the numerator or denominator is consistent with zero.
The ratios for the combined spectra in the inner, polar bin are statistically different from  those in any other radial or azimuthal bin (confidence between 97 and 99.9\% depending on the test). 

Our identification of azimuthal variations in the mean CGM properties is qualitatively consistent with the \cite{Ignacio2021} conjecture, but does not necessarily confirm it. We are limited by the statistical confidence of our results and await larger, deeper spectroscopic samples with which to revisit these results. There are prospects in this regard.
The DESI \citep[Dark Energy Spectroscopic Instrument,][]{DESI1, DESI2}  will ultimately collect $\sim$ thirty million spectra,  an increase in the number of sightlines  by two orders of magnitude relative to the current study due both to the increase in the number of sightlines and the number of suitable primaries. Aside from the gain in numbers, those spectra should also be better suited for our purpose than SDSS spectra because DESI targeted fainter objects and reach, on average, a lower S/N for each `nuisance' target, thereby making continuum subtraction easier and less critical.
Furthermore, proposed massively multiplexed deep spectroscopic surveys with new, dedicated telescopes will push even further. Although we eagerly await the mapping of emission lines in individual galaxies with which to explore the CGM, there is still a role for these stacked analyses to examine mean properties and address a wide array of scientific questions.

\section{Acknowledgments}

HZ acknowledges financial support from the start-up funding of the Huazhong University of Science and Technology. DZ acknowledges financial support from NSF grant AST-2006785. The authors gratefully acknowledge the SDSS III team for providing a valuable resource to the community.
Funding for SDSS-III has been provided by the Alfred P. Sloan Foundation, the Participating I institutions, the National Science Foundation, and the U.S. Department of Energy Office of Science. The SDSS-III web site is http://www.sdss3.org/.

SDSS-III is managed by the Astrophysical Research Consortium for the Participating Institutions of the SDSS-III Collaboration including the University of Arizona, the Brazilian Participation Group, Brookhaven National Laboratory, Carnegie Mellon University, University of Florida, the French Participation Group, the German Participation Group, Harvard University, the Instituto de Astrofisica de Canarias, the Michigan State/Notre Dame/JINA Participation Group, Johns Hopkins University, Lawrence Berkeley National Laboratory, Max Planck Institute for Astrophysics, Max Planck Institute for Extraterrestrial Physics, New Mexico State University, New York University, Ohio State University, Pennsylvania State University, University of Portsmouth, Princeton University, the Spanish Participation Group, University of Tokyo, University of Utah, Vanderbilt University, University of Virginia, University of Washington, and Yale University.

\bibliography{bibliography}

\end{document}